\documentclass{article}
\usepackage{spconf,amsmath,graphicx,amsfonts}

\title{Multiresolution Signal Processing of \\ Financial Market Objects}
\name{Ioana Boier}
\address{NVIDIA \\
  iboier@nvidia.com
}
\begin{document}
%
\maketitle
\begin{abstract}
Multiresolution analysis has applications across many disciplines in the
study of complex systems and their dynamics. Financial markets are
among the most complex entities in our
environment, yet mainstream quantitative models operate at
predetermined scale, rely on linear correlation measures, and
struggle to recognize non-linear or causal structures. In this paper,
we combine neural networks known to capture non-linear associations
with a multiscale decomposition to facilitate a better
understanding of financial market data substructures. Quantization
keeps our decompositions calibrated to market at every
scale. We illustrate our approach in the context of seven use cases.

\end{abstract}
\section{Introduction}
\label{intro}

Financial markets are prototypical examples of systems exhibiting
multiple scales of behavior. Micro and macro-economic factors
intertwine in surreptitious ways, driving the ups and downs of
investment portfolios. Signal processing (SP) techniques have
been adapted from engineering to finance
in search of alpha, or systematic patterns that emerge from data
leading to excess returns.
Their main challenge lies in dealing with the high dimensionality of the
data \cite{akansu2016financial}.
Machine learning (ML) applied to finance uses pattern
learning paradigms closely connected to traditional
statistical and numerical approaches
\cite{akansu2016financial}. ML has its own
challenges: financial data is non-stationary, noisy, and often
insufficient, given the high dimensionality of the space to which it belongs.

\begin{figure}[htb]

\begin{minipage}[b]{1.0\linewidth}
  \centering
  \centerline{\includegraphics[width=8.5cm]{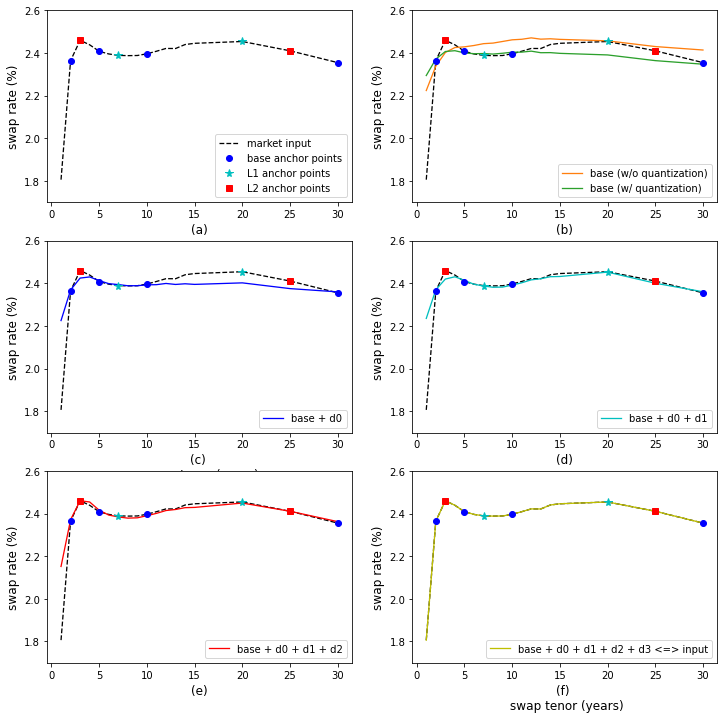}}
\end{minipage}
\caption{Multiresolution decomposition of the swap curve of March 21,
  $2022$. Anchor points are assigned to various scales by importance
  (e.g., market liquidity). The input curve (a) is decomposed into
  approximations from coarse to fine, calibrated to the anchors on
  each layer (base and $L_0$ layers share the same
  anchors). Figures (b) through (e) illustrate the intermediate
  outputs from our FinQ-VAE model, while (f), reproduces the input after
  incorporating the final residuals.}
\label{fig:mres}
\end{figure}

One way to reduce dimensionality is to consider the intrinsic
structure present in the data. For instance, bond yields, swap rates,
inflation, and foreign exchange (FX) rates can be thought of as having
$1$D term-structures (e.g.,
zero-coupon, spot, forward, or basis curves).  Similarly, volatilities
implied by option prices are organized as (hyper-)surfaces in $2$D or
higher. These structures carry information that can help reduce
complexity.
A typical spot swap curve could have as many as 
fifty tenors. Studying the corresponding time series
means working in $50$D space. A $10 \times
10$ volatility grid puts us in $100$D space. Knowing that these data
lie, in fact, on $n$D manifolds where $n$ is relatively small, greatly
reduces the burden of the learning task. We refer to these
market data structures as {\it market objects} and focus on learning market behaviors from the shape and
dynamics of these representations.

We propose a {\it multiresolution
  decomposition} of market objects generated with a novel 
architecture (FinQ-VAE) consisting of a pipeline of variational
autoencoders (VAE)
with latent space quantizations guided by
{\it financially meaningful constraints}, e.g., market liquidity or
trading views. Figure~\ref{fig:mres} shows a learned multiresolution
decomposition of the US swap curve of March $21$, $2022$.

\section{Related Work}
\label{relwork}

Financial data, i.e., time series of prices, are typically
{\it non-stationary} \cite{cont2001empirical}.
Hence, most SP and ML
approaches operate on {\it returns}, i.e., changes in price from one time stamp to the
next. A time series of returns is typically more ``stable'' and likely
to pass stationarity tests \cite{tsay2005analysis}. However,
financial data also suffers from {\it low signal-to-noise
  ratios}. Hence, any signal 
found is likely wiped out by differencing.
To quote L\'opez de Prado \cite{de2018advances}, ``returns are
stationary, however memoryless, and prices have memory, however they
are non-stationary''.

Multiresolution methods can be traced all the way back to Fourier
transforms and wavelets \cite{haar1909theorie,
  daubechies1988time, mallat1989multiresolution}. They balance 
signal preservation with stationarity:
a base shape acts as a noise smoother,
retaining an average signal level, and a hierarchy of residuals
adds refinements to the base shape while exhibiting desirable
statistical properties like stationarity.
In finance, multiscale approaches
have primarily focused on scaling along the time dimension
(minute-by-minute, daily, weekly, etc.).

We propose a novel way of learning multiresolution decompositions of
financially meaningful sub-structures in the data.
The inspiration comes
from multiresolution geometry processing techniques for fairing and
interactive editing \cite{boier2005detail}. The analogy with finance
lies not only in the need to disentangle or denoise a fair shape from
a noisy representation, but also in the need for scenario generation through controlled movements of {\it key
  points}. In modeling physical interactions, the type
of material dictates the influence of a single edit on the surrounding
shape. In modeling financial scenarios, the influence of certain
points on the dynamics of their neighbors is subject to financial
constraints.
Points on a yield curve or on a volatility surface don't move in
isolation.
They are correlated to points around them.
The region of influence of such move depends on the use case.
Some moves
have far-reaching implications, others are more localized.
The nature of the deformation is also constrained by laws of
arbitrage and financial plausibility.

Autoencoders with their variational and conditional flavors
\cite{rumelhart1985learning, kingma2013auto, sohn2015learning} have
been adopted in finance mostly for single-resolution latent learning
and its applications \cite{kondratyev2018learning,
  bergeron2022variational, suimon2020autoencoder, sokol2021aemm,
  lim2020detecting, gu2021autoencoder}.
We extend these ideas in two significant ways: (a)~we define a new
architecture of cascading VAEs to learn hierarchical decompositions
of market objects and we illustrate how to leverage them in a variety
of applications and (b) we introduce a quantization
step \cite{razavi2019generating} that takes into account financially
meaningful constraints to ensure calibration to market at every scale. 

\section{FinQ-VAE}
\label{finq}

\subsection{Modeling Background}
VAEs facilitate the learning of probabilistic generative models from a
set of observations in $x \in \mathbb{R}^d$ presumed to lie on a
manifold of dimension $r \leq d$. The loss function to be optimized
during training is according to \cite{higgins2017betavae}:

\begin{center}
$\mathcal{L}(\theta, \phi, \beta; x, z) = \mathbb{E}_{q_{\phi}(z|x)}[log p_{\theta}(x|z)] - \beta D_{KL}(log q_{\phi}(z|x) || p(z)) $
\end{center}

With its roots in
signal processing theory, {\it quantization} helps compress continuous
sets of values to a countable or even finite number of options beyond
which the variability in values makes little or no contribution to
modeling. For example, images can be viewed as containing
discrete objects along with a discrete set of
qualifiers such as color, shape, texture.
Vector-quantized VAEs (VQ-VAE) have been developed to support discrete
learning \cite{razavi2019generating, van2017neural}.
Market objects also contain redundant information. Moreover, like higher-level objects in images,
they are, by definition, discrete collections of
sub-objects with qualifiers such as steepness, curvature,
skews, smiles, liquid and illiquid regions.
Therefore, it makes sense to consider the benefits of quantizing their latent spaces.
By default, however, VAE outputs are not faithful to specific data.
In finance it is important to calibrate models to {\it
  market-observed data} or to {\it enforce desired
  constraints}. Instead of using a full-fledged VQ-VAE, we employ a
simpler, non-learned quantization process that snaps encoder outputs
to optimal latent locations via optimization with constraints. These
dynamically quantized points are passed to the decoder to produce
calibrated reconstructions which are used to compute residuals
for the next layer. Consequently, our multiresolution
decompositions are {\it calibrated to market at each scale}.

\subsection{Problem Statement}

Given a market object $\mathcal{O}_{mkt}$, our goal is to learn a
multiresolution decomposition such that:

\begin{enumerate}
\item It consists of a base shape object $\mathcal{O}_{base}$ and a number $n+1$ of residual layers, $L_0, L_1, ..., L_{n}$ and objects $\delta_0, \delta_1, ..., \delta_{n}$ such that $\mathcal{O}_{mkt} = \mathcal{O}_{base} + \delta_0 + \delta_1+ ... +\delta_{n}$.
\item {\it Anchor points} are selected on each of the base and
  residual layers $L_0, L_1, ..., L_{n-1}$: $a_i^j$, where $i$ is the
  index of the anchor and $j$ is the layer index. The selection
  criteria reflect financial considerations: e.g., some points are
  more liquid or tradeable.
\item Each intermediate reconstruction with $j < n$ layers $\mathcal{O}_{j} = \mathcal{O}_{base} + \delta_0 + \delta_1+ ... +\delta_{j-1}$ is respectively calibrated to anchor points on $L_0, L_1, ..., L_{j-1}$. The residuals in the last layer $L_{n}$ are computed to recover the input market object exactly: $\delta_{n} = \mathcal{O}_{mkt} - \mathcal{O}_{n-1} $.
  
\end{enumerate}

\begin{figure}[htb]

\begin{minipage}[b]{1.0\linewidth}
  \centering
  \centerline{\includegraphics[width=8cm]{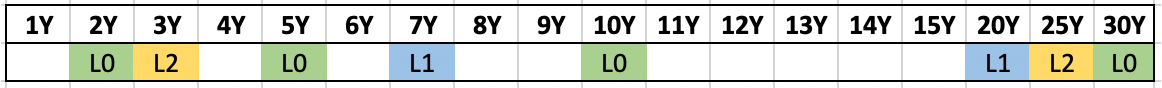}}
   \centerline{(a) Swap curve}\medskip
\end{minipage}
\begin{minipage}[b]{1.0\linewidth}
  \centering
  \centerline{\includegraphics[width=8.5cm]{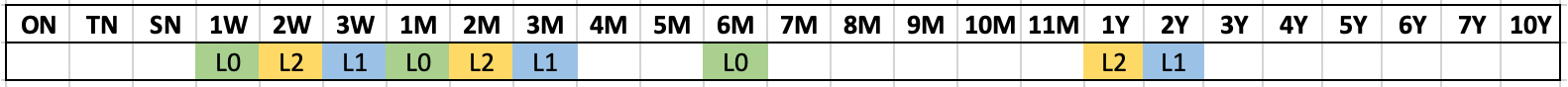}}
  \centerline{(b) FX forward curve}\medskip
\end{minipage}
\begin{minipage}[b]{1.0\linewidth}
  \centering
  \centerline{\includegraphics[width=5.5cm]{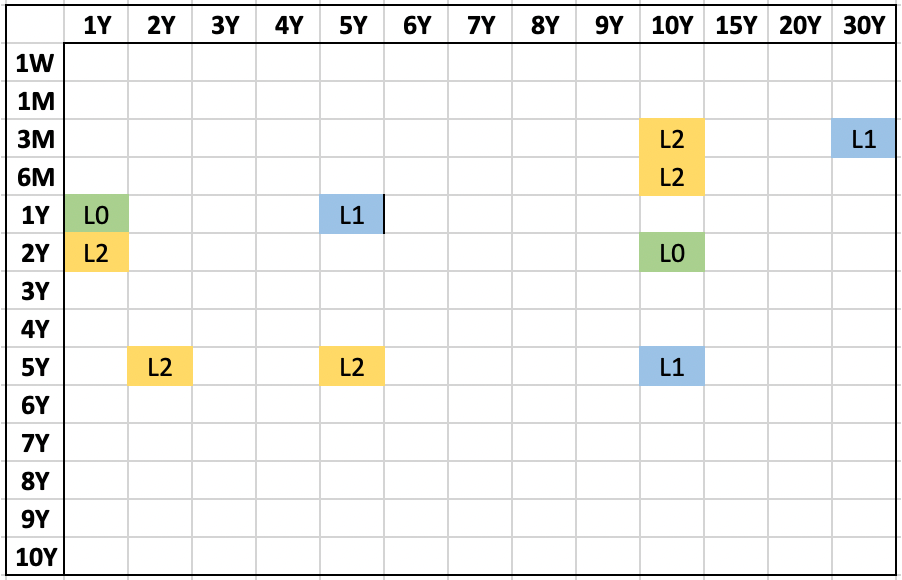}}
  \centerline{(b) Swaption volatility surface}\medskip
\end{minipage}

\caption{User-specified anchor points for (a) a swap curve, (b) an FX
  forward curve, (c) a swaption volatility surface.}
\label{fig:mres_layers}
\end{figure}

Figure \ref{fig:mres_layers} shows examples of user-defined
anchors distributed according to market
liquidity. We note that this technique 
applies to market objects of different dimensions with scattered
constraints that need not be regularly spaced.

\subsection{The FinQ-VAE Architecture}

\begin{figure}[htb]

\begin{minipage}[b]{1.0\linewidth}
  \centering
  \centerline{\includegraphics[width=8.5cm]{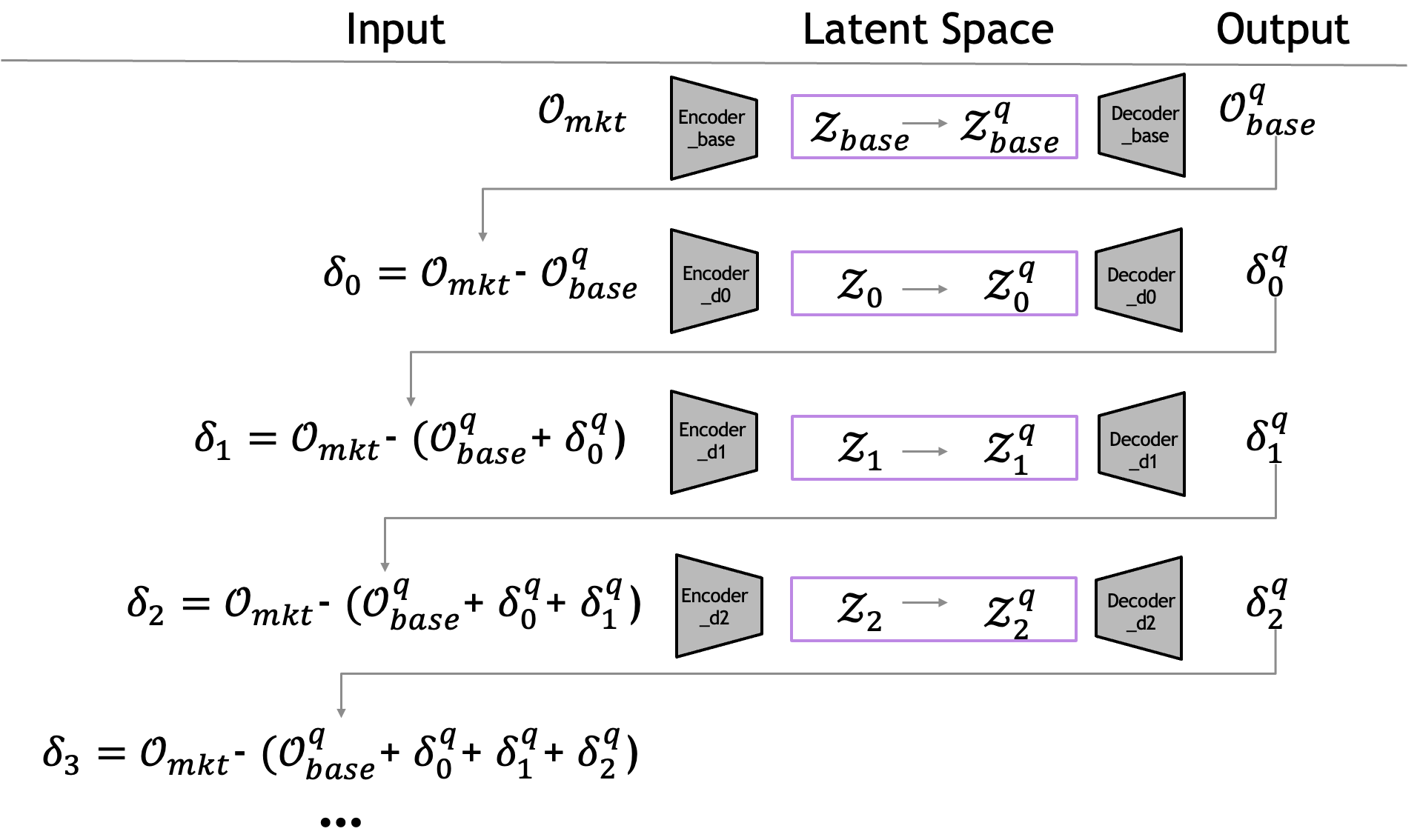}}
\end{minipage}
\caption{FinQ-VAE: a pipeline of VAEs with financially quantized
  latent spaces is trained to learn a multiresolution decomposition of
  a market object into a base shape and several layers of residuals.}
\label{fig:arch}
\end{figure}

Our FinQ-VAE architecture is shown in Figure \ref{fig:arch}.
Without loss of generality, we illustrate it in the context of swap curves, but the same concept applies to other types of market objects.
The training inputs are market objects
$\mathcal{O}_{mkt}$ and anchors $\mathcal{A}$ (Figure
\ref{fig:mres} (a)).
Each layer is a VAE neural net, consisting of an
Encoder, a Decoder, and a latent space that entails a quantization
that maps encoded latent vectors into constraint-optimized
vectors to be passed to the Decoder for reconstruction. 

\begin{figure}[htb]

\begin{minipage}[b]{1.0\linewidth}
  \centering
  \centerline{\includegraphics[width=8.5cm]{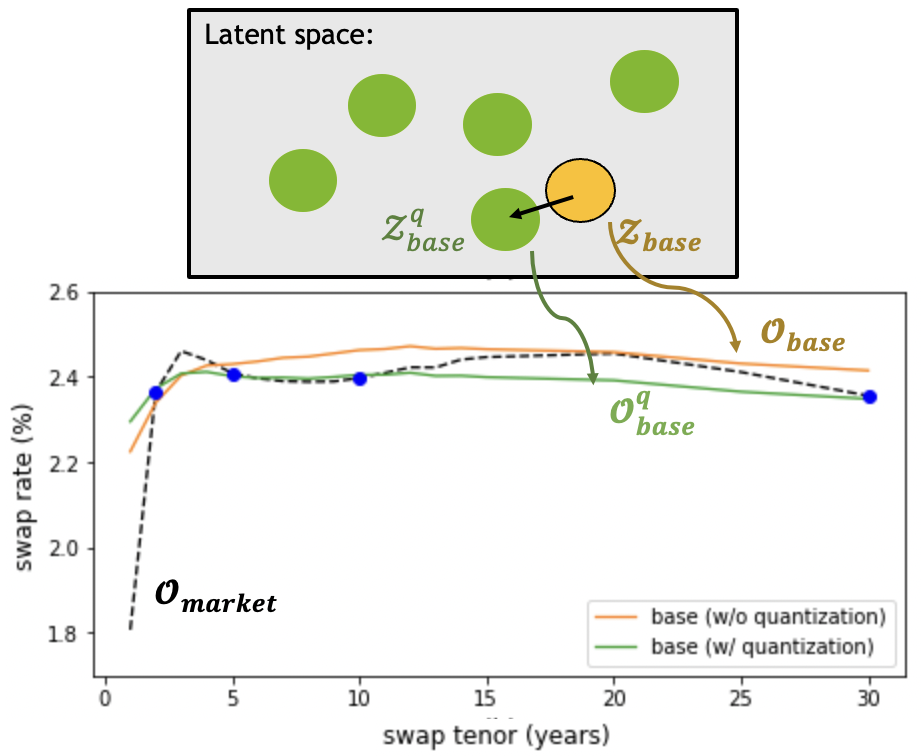}}
\end{minipage}
\caption{Base curve reconstruction with constraints at anchor points (blue dots), before (orange curve) and after (green curve) quantization in latent space.}

\label{fig:qproc}
\end{figure}

The {\it base layer} learns a coarse general shape (Figure
\ref{fig:qproc}). In this example, the $Encoder_{base}$ takes as input
the full swap curve specified by a set of $m=18$ swap tenors and a set
of base anchor points, e.g., $\mathcal{A} = \{2$Y, $5$Y, $10$Y,
  $30$Y$\}$ (Figure \ref{fig:mres_layers}). The output of 
$Encoder_{base}$  is a latent vector $z_{base}$ in a latent space of
fixed dimension.
In this example we used a $3$D latent space. Without quantization,
$Decoder_{base}$ maps the latent vector $z_{base}$ into a {\it base}
output curve $\mathcal{O}_{base}$, i.e., a reconstruction representing
the ``learned'' global shape.

This representation can be likened to a PCA reconstruction using the same number of principal components. Unlike a PCA
output, the VAE-generated curves are better fitted to anchors by
design. We have modified the VAE loss function to include an anchor calibration term:

\begin{center}
  $\mathcal{L}_{recon}(x, Decoder(z)) = ||x-Decoder(z)||_2^2+\alpha ||x_{|\mathcal{A}}-Decoder(z)_{|\mathcal{A}}||_2^2$
\end{center}

\noindent
where $\alpha \geq 0$ and the anchors $\mathcal{A}$ serve as
constraints.

\begin{figure}[htb]

\begin{minipage}[b]{1.0\linewidth}
  \centering
  \centerline{\includegraphics[width=8.5cm]{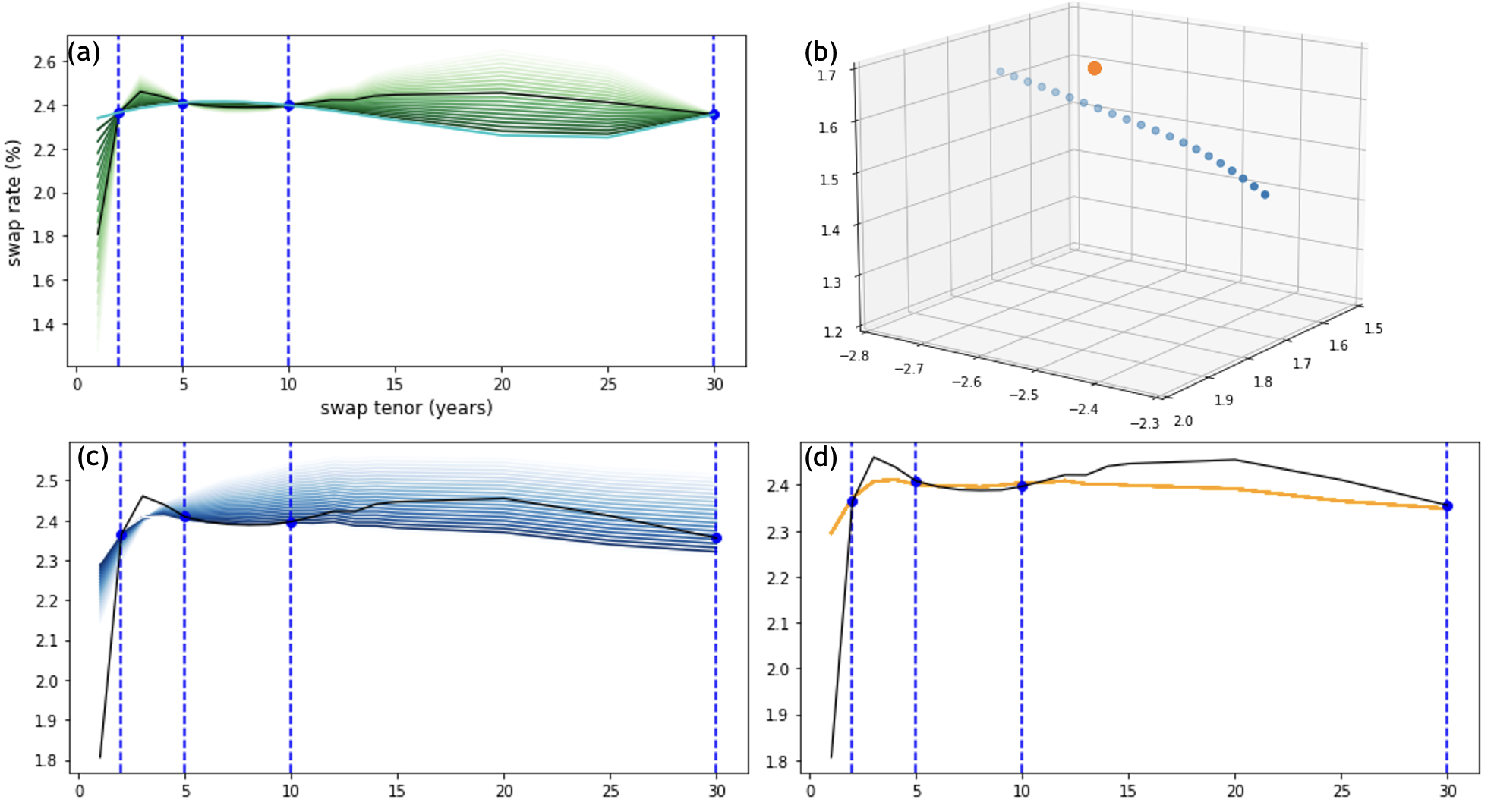}}
\end{minipage}
\caption{Reconstructions with and without quantization: (a) input curves
  pass through the same $L_0$ anchors; (b) embeddings of the base
  shapes in $3$D latent space:  blue points are embeddings before
  quantization, orange points (overlapping) are embeddings after
  quantization; (c) base curve reconstructions without quantization are
  not well calibrated to the anchors; (d) a well-fitted base shape is
  obtained after quantization.}

\label{fig:quanti}
\end{figure}

While the reconstructed base curve fits the overall shape, we would
like the anchor points to be fitted even better.
We quantize the latent
vector $z_{base}$ to $z_{base}^q$ via optimization:

\begin{equation}
  z_{j}^q = argmin_z ||x_{|\mathcal{A}^j}-Decoder_{j}(z)_{|\mathcal{A}^j}||_2^2  
\end{equation}
  
\noindent
where $j$ is the layer index. The base curve $\mathcal{O}_{base}^q$
reconstructed from the quantized vector is shown in Figure
\ref{fig:mres} (b). Figure \ref{fig:quanti} shows curves with the same
base shape before and after quantization.
Given a set of artificially generated input curves passing through the same anchors in (a), the corresponding embeddings of their base shapes in $3$D latent space are shown in (b):  blue points are the embeddings before quantization, orange points (overlapping) are the embeddings after quantization. The base curve reconstructions without quantization are not well calibrated to the anchors as shown in (c). Reconstructions with quantization produce a well-fitted base shape in (d).

The first {\it residual layer} takes as input the difference object
$\delta_0 = \mathcal{O}_{mkt} - \mathcal{O}_{base}^q$ and another VAE
is trained to learn the shape of $\delta_0$ residuals. The resulting quantized
residual $\delta_0^q$ is applied to the
base object $\mathcal{O}_{base}^q$ to produce our $L_0$ market object
reconstruction: $\mathcal{O}_0 = \mathcal{O}_{base}^q+\delta_0^q$, see Figure \ref{fig:mres} (c).

The second residual layer takes as input the difference object
$\delta_1 = \mathcal{O}_{mkt} - \mathcal{O}_0$ and the process is
repeated. Figures \ref{fig:mres} (d)-(e) show the reconstructions on
layers $1$ and $2$. The final residuals $\delta_3 = \mathcal{O}_{mkt}-\mathcal{O}_{2}$ and computed to recover the input $\mathcal{O}_{mkt}$ exactly (Figure \ref{fig:mres} (f)).

\vspace{-0.3cm}
\section{Applications and Results}
\label{results}

\begin{figure}[htb]
\begin{minipage}[b]{1.0\linewidth}
  \centering
  \centerline{\includegraphics[width=8cm]{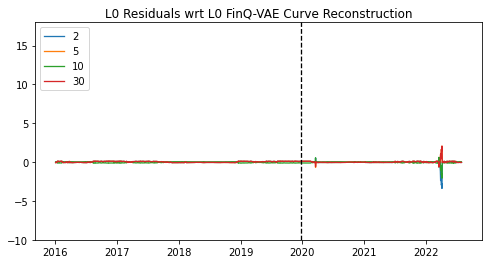}}
  \centerline{(a)}\medskip
\end{minipage}
\begin{minipage}[b]{1.0\linewidth}
  \centering
  \centerline{\includegraphics[width=8cm]{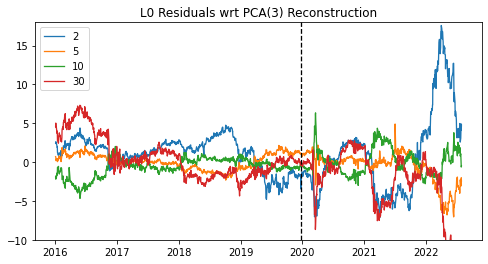}}
  \centerline{(b)}\medskip
\end{minipage}
\caption{(a) Liquid point residuals ($2$Y, $5$Y, $10$Y, $30$Y) with
  respect to the $L_0$ FinQ-VAE reconstruction are
  relatively small in magnitude and stable, including over the test period
  starting in $2020$. (b) Shown comparatively are the residuals with
  respect to a PCA reconstruction with $3$ factors.}
\label{fig:resid2y}
\end{figure}

In this section we present seven use cases.
Objects reconstructed with FinQ are not only plausible
as in \cite{henry2019generative}, but
also calibrated to market.
The choice of anchors at different scales is fully configurable.
We trained two FinQ models: one on daily USD
spot swap curves and one on bond curves \cite{fed} between Jan $2001$
and the end of $2019$. The test data period is Jan $2020$ to Jul
$2022$.
Using the magnitude of the $L_0$ residuals as a measure of the quality
of fit between the learned base curves and the actual market data around
the most liquid points, we note that 
the model appears robust over the test data, starting in 2020. This includes 
the COVID-19 pandemic, the inflationary period that follows, and the
start of the rate hike cycle by the
US Fed, as evident in Figure~\ref{fig:resid2y}.

\vspace{-0.3cm}
\subsubsection*{Hierarchical Factor Analysis}

\begin{figure}[htb]

\begin{minipage}[b]{.48\linewidth}
  \centering
  \centerline{\includegraphics[width=4.5cm]{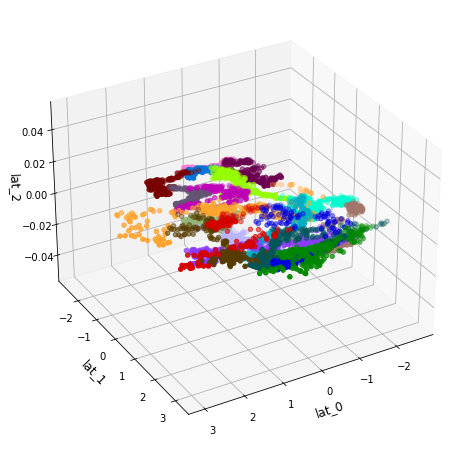}}
  \centerline{(a) Base shape ($3$D)}\medskip
\end{minipage}
\hfill
\begin{minipage}[b]{0.48\linewidth}
  \centering
  \centerline{\includegraphics[width=4.5cm]{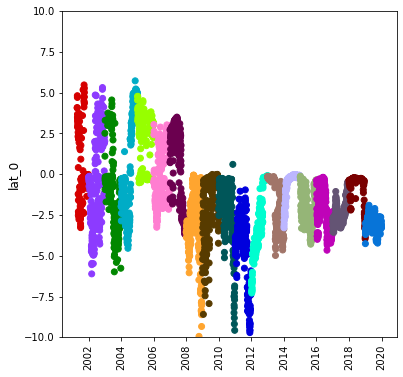}}
  \centerline{(b) $L_0$ residuals ($1$D)}\medskip
\end{minipage}
\begin{minipage}[b]{.48\linewidth}
  \centering
  \centerline{\includegraphics[width=4.5cm]{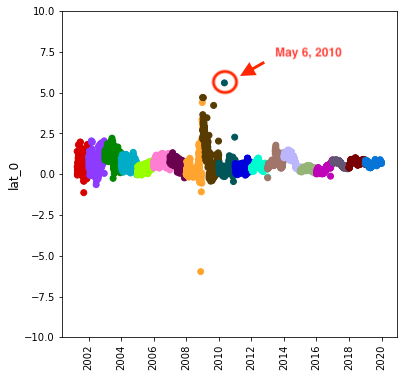}}
  \centerline{(c) $L_1$ residuals ($1$D) w/ outlier}\medskip
\end{minipage}
\hfill
\begin{minipage}[b]{0.48\linewidth}
  \centering
  \centerline{\includegraphics[width=4.5cm]{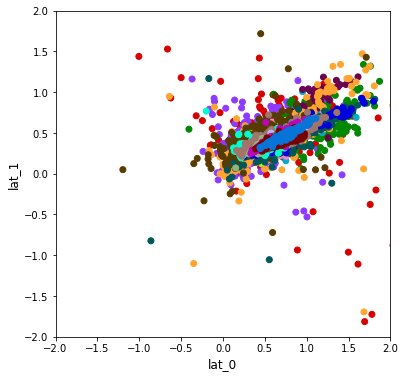}}
  \centerline{(d) $L_2$ residuals ($2$D)}\medskip
\end{minipage}
\caption{Hierarchy of latent spaces (color-coded by year).}
\label{fig:hier_lat}
\end{figure}

In contrast to PCA which typically operates on returns and ignores
market levels, our multiresolution decomposition learns a global shape
level on its base layer and residuals at various scales.
The variational feature of the model ensures smooth
navigation through latent space. The quantization feature ensures that
outputs are calibrated to user-specified important points. The latter
is a powerful property, as traditionally, VAEs have only been used in
finance for generating ``similar'' data to some learned distribution
that could be rather different from market-observed values.
Figure \ref{fig:hier_lat} illustrates the hierarchy of latent spaces
in our learned model.

\begin{figure}[htb]

\begin{minipage}[b]{1.0\linewidth}
  \centering
  \centerline{\includegraphics[width=8.5cm]{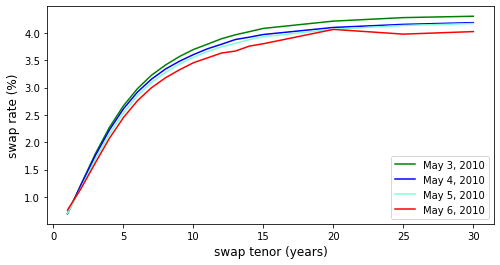}}
\end{minipage}
\caption{Outlier detection: a closer inspection around May 6, $2010$
  reveals a curve with noise around the $20$Y tenor which is an anchor
  on the $L_1$ level. The resulting non-smooth curve does not
  correspond to the learned shape distribution, hence it appears as an
  outlier in the $L_1$ latent space (see Figure \ref{fig:hier_lat} (c)).}
\label{fig:out}
\end{figure}

\vspace{-0.3cm}
\subsubsection*{Scenario Generation}

Standard practice in financial risk management is to shock market
objects either in absolute or percentage terms. There are two main
considerations: the shape of a scenario and its overall
size.
Traditional techniques to
generate plausible scenarios revolve around historical
deformations and/or artificial stresses. Unfortunately, history
doesn't often
repeat itself and the generation of artificial scenarios is
rather empirical: one may have a view of the movement in certain
regions of the market, but the dependence of the rest of
the market to movements in those regions may be difficult to
ascertain.
PCA-based techniques are appealing because of their
simplicity, however proper calibration of scenario size
is challenging and global dependence on non-intuitively weighted
linear combinations of points is difficult to
interpret. 

\begin{figure}[htb]

\begin{minipage}[b]{1.0\linewidth}
  \centering
  \centerline{\includegraphics[width=8cm]{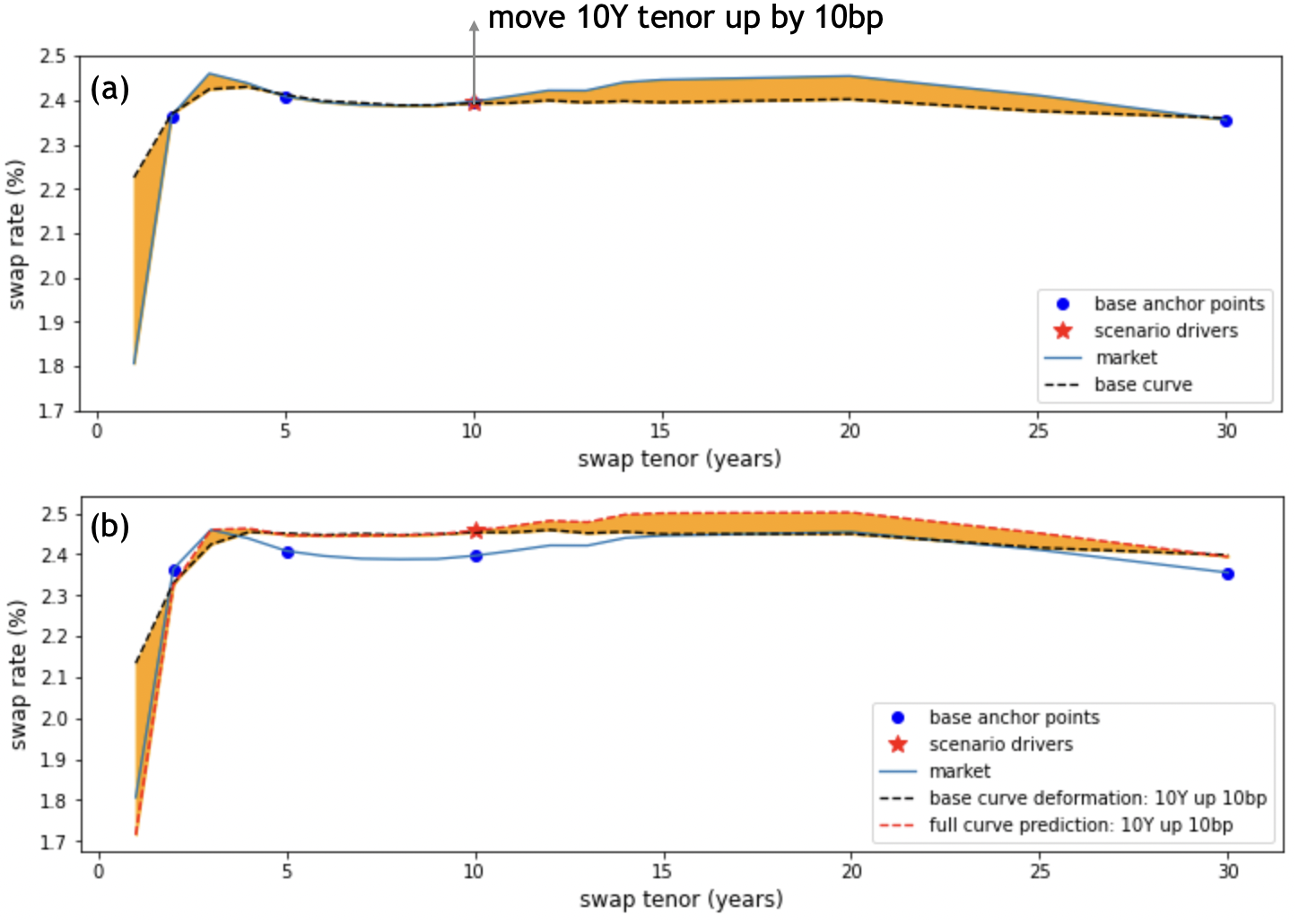}}
\centerline{(a) $10$Y point moves up by $10$bp.}\medskip
\end{minipage}
\begin{minipage}[b]{1.0\linewidth}
  \centering
  \centerline{\includegraphics[width=8cm]{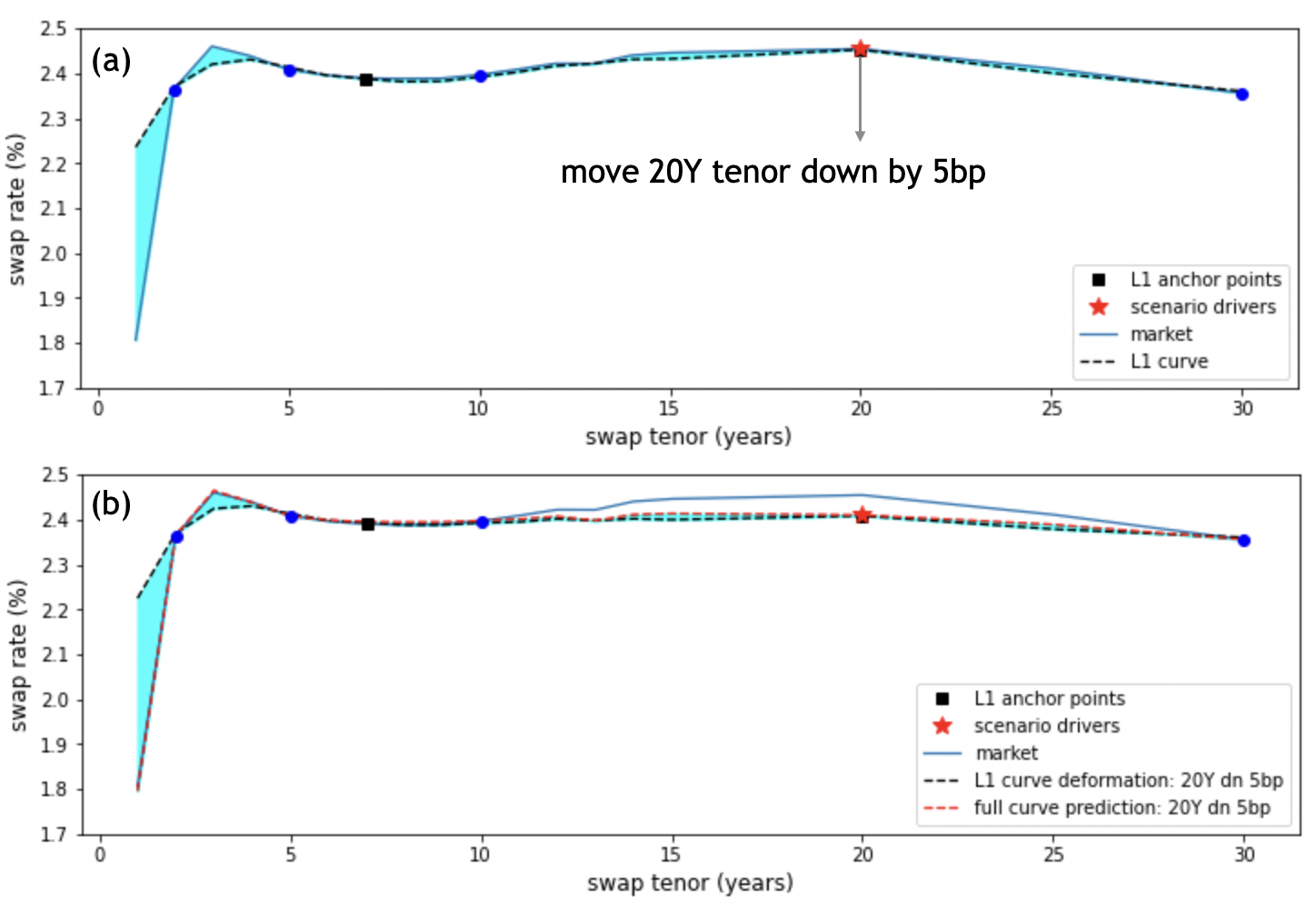}}
  \centerline{(b) $20$Y point moves down by $5$bp.}\medskip
\end{minipage}

\begin{minipage}[b]{1.0\linewidth}
  \centering
  \centerline{\includegraphics[width=8cm]{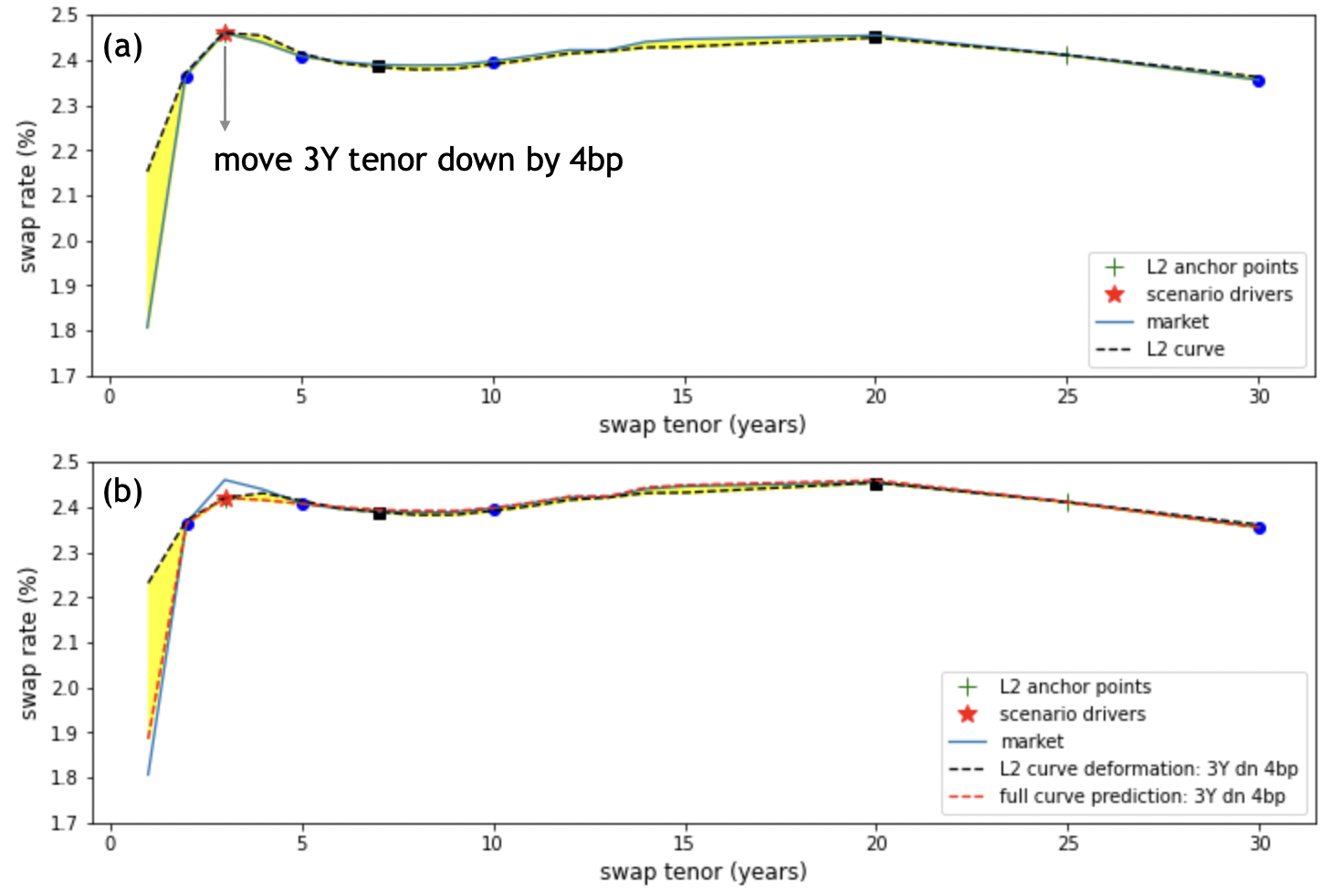}}
  \centerline{(c) $3$Y point moves down by $4$bp.}\medskip
\end{minipage}
\caption{Full curve scenarios conditional on user moves. Coarse-to-fine anchor points have global-to-local impacts.
}
\label{fig:scen}
\end{figure}

Our multiresolution framework splits
responsibilities: base shapes account for market levels and are under
the control of a few key drivers that are easier to intuit by human
experts. Their views define anchor points and desired
stresses. Dependencies that are more difficult to synthesize through human
experience are generated algorithmically
(Figure \ref{fig:scen}).

\vspace{-0.3cm}
\subsubsection*{Synthetic Data Generation}

\begin{figure}[htb]

\begin{minipage}[b]{.48\linewidth}
  \centering
  \centerline{\includegraphics[width=4.5cm]{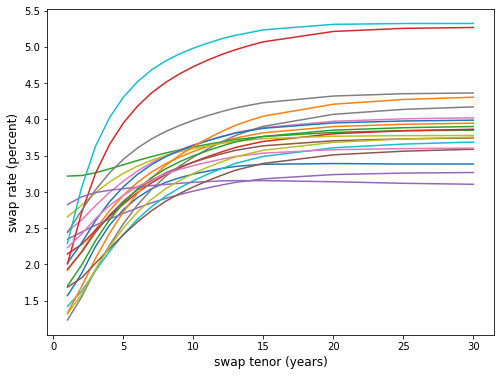}}
  \centerline{(a) base curves}\medskip
\end{minipage}
\hfill
\begin{minipage}[b]{0.48\linewidth}
  \centering
  \centerline{\includegraphics[width=4.5cm]{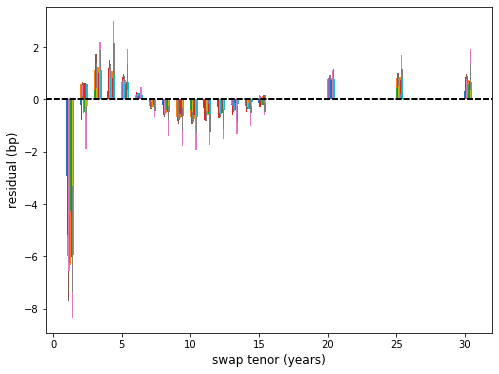}}
  \centerline{(b) $L_0$ residuals}\medskip
\end{minipage}
\begin{minipage}[b]{.48\linewidth}
  \centering
  \centerline{\includegraphics[width=4.5cm]{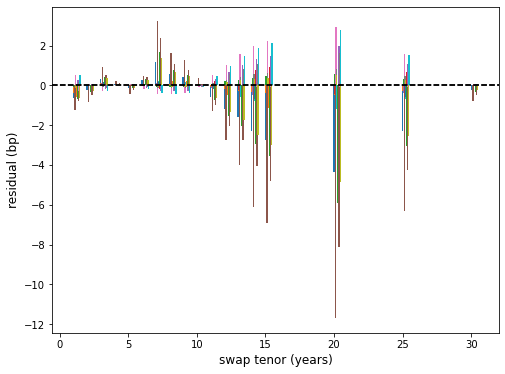}}
  \centerline{(c) $L_1$ residuals}\medskip
\end{minipage}
\hfill
\begin{minipage}[b]{0.48\linewidth}
  \centering
  \centerline{\includegraphics[width=4.5cm]{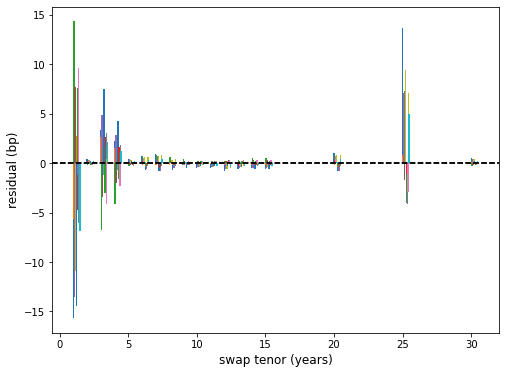}}
  \centerline{(d) $L_2$ residuals}\medskip
\end{minipage}
\caption{Synthetic samples generated at various scales.}
\label{fig:sdg}
\end{figure}

Synthetic market objects are composable in hierarchical fashion. This
can be done artificially by sampling the latent spaces of
the VAEs from coarse to fine, or by using historical or
trader-specified moves of anchor points for the lower layers in the
hierarchy and randomly sampling the higher ones. Figure \ref{fig:sdg} illustrates synthetic samples 
on each level.

\vspace{-0.3cm}
\subsubsection*{Nowcasting}

Nowcasting is the ``forecasting" of the present or the near future in
the absence of complete information about the current state of the
market. It has two main components: an
understanding of what is already priced in the market and a view on
future conditions. We allow for such views to be
incorporated. For example, option portfolios
require full implied volatility surfaces to price, yet 
option prices may not be available for all $(expiry,
underlier)$ pairs. The most liquid points can be used
as anchors, while missing values can be sampled from
the latent distributions. In the examples of Figure~\ref{fig:scen},
the scenario curves are reconstructed from the known current move of a single
point (scenario driver) and previous residuals. Unlike conditional models, our
approach does not require training with conditional labels.

\vspace{-0.3cm}
\subsubsection*{Residuals as Signals}

Residual time series could be used as signals
for systematic strategies. $\mathcal{A}_0$ anchor points should
have residuals that are close to zero on $L_0, L_1, \cdots$. If this is not the case, their deviation
from zero may be used as a signal that the shapes being reconstructed
are difficult to fit to the constraints. Such difficulties are
harbingers of unusual market conditions.

\begin{figure}[htb]
\begin{minipage}[b]{1.0\linewidth}
  \centering
  \centerline{\includegraphics[width=8cm]{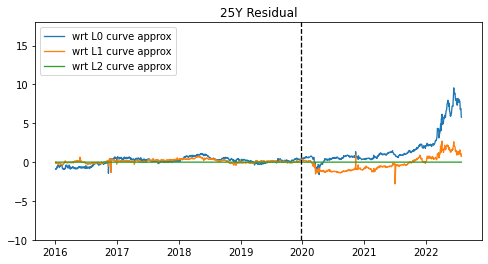}}
\end{minipage}
\caption{$25$Y residuals vs. $L_0, L_1, L_2$ reconstructions.}
\label{fig:resid}
\end{figure}

Figure \ref{fig:resid}
depicts residuals of the $25$Y swap rate with respect to the
FinQ-generated $L_0, L_1, L_2$ curves (vertical line is the boundary
between train/test data). The residuals are small
with respect to the $L_2$ reconstruction since $25$Y is an anchor on
$L_2$. They are larger, but mean reverting vs. coarser
reconstructions $L_1$ and $L_0$, respectively.

\vspace{-0.3cm}
\subsubsection*{Relative Value Analysis}

\begin{figure}[htb]

\begin{minipage}[b]{1.0\linewidth}
  \centering
  \centerline{\includegraphics[width=8cm]{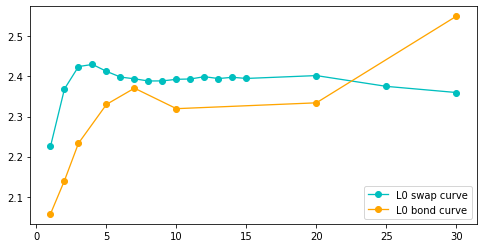}}
\end{minipage}
\caption{Relative value analysis of bond yields vs. swap rates.}
\label{fig:relval}
\end{figure}

We applied FinQ to learning US
Treasury bond curves. While bonds and interest rate swaps capture similar
macroeconomic developments, swap curves tend to be smoother. Hence, swap curve reconstructions could be used to
identify viable swap spread trades. Figure \ref{fig:relval} indicates
that buying $30$Y bonds and selling the same maturity swaps might
be a good strategy.

\vspace{-0.3cm}
\subsubsection*{Outlier Detection}

The variational aspect of autoencoders ensures relatively compact clusters of latent
encodings. Samples that stand out from their clusters are likely
to be outliers. Figure \ref{fig:hier_lat} (c) singles out
such a point corresponding to May 6, $2010$; upon closer inspection of
the data (Figure~\ref{fig:out}),
we notice that on this date the $20$Y value seems stale, causing a
non-smooth curve (the ``flash
crash'' occurred intraday on that date, which may have contributed to noise in the
end-of-day data).

\section{Conclusions}
\label{conclusions}

FinQ-VAE is a novel architecture for
multiresolution signal processing of market objects. 
Market-calibrated representations are learned using a layered approach. User-specified constraints can be incorporated at different
scales to generate quantized embeddings that lead to
calibrated reconstructions of market objects. To our knowledge, this
is the first time multiresolution analysis is combined with quantized
VAEs and applied to financial modeling in a way that accommodates
constraints such as trading views, liquidity,
etc.
We showed that the resulting decompositions could serve in a
variety of use cases.
Our technique applies across asset classes and different dimensionality structures.  


\bibliographystyle{IEEEbib}
\bibliography{boier_finq_arxiv2}

\end{document}